\documentclass{aa}
\usepackage{graphicx}
\usepackage{txfonts}
\usepackage{color}
\usepackage{longtable,lscape}
\usepackage[]{natbib}

\begin{document}
   \title{Star formation in isolated AMIGA galaxies: dynamical influence of bars\thanks{Partially based on observations collected at the Centro Astron\'omico Hispano Alem\'an (CAHA) at Calar Alto, operated jointly by the Max-Planck Institut f\"ur Astronomie and the Instituto de Astrof\'isica de Andaluc\'ia (CSIC), as well as at the Observatorio de Sierra Nevada (OSN), operated by the IAA/CSIC.}}

   \subtitle{}

   \author{S. Verley    \inst{1,2,3}
        \and
        F. Combes       \inst{1}
        \and
        L. Verdes-Montenegro    \inst{2}
        \and
        G. Bergond      \inst{2,4}
        \and
        S. Leon \inst{5,2}
          }


   \institute{LERMA, Observatoire de Paris, 61, avenue de l'Observatoire, 75014 Paris, France\\
              \email{Simon.Verley@obspm.fr}
        \and
             Instituto de Astrof\'isica de Andaluc\'ia -- CSIC, C/ Camino Bajo de Hu\'etor, 50, 18008 Granada, Spain
        \and
INAF, Osservatorio Astrofisico di Arcetri, Largo E. Fermi, 5, 50125 Firenze, Italy
        \and
GEPI, Observatoire de Paris-Meudon, 61, avenue de l'Observatoire, 75014 Paris, France
        \and
             Instituto de RadioAstronom\'ia Milim\'etrica, 
             Avenida Divina Pastora, 7, N\'ucleo Central, 18012 Granada, Spain\\
             }

   \date{Received Month DD YYYY / Accepted Month DD YYYY}

 
  \abstract
   {Star formation depends strongly both on the local environment of galaxies, and on the internal dynamics of the interstellar medium. To disentangle the two effects, we obtained, in the framework of the AMIGA project, H$_\alpha$ and Gunn r photometric data for more than 200 spiral galaxies lying in very low-density regions of the local Universe.}
   {We characterise the H$_\alpha$ emission, tracing current star formation, of the 45 largest and less inclined galaxies observed for which we estimate the torques between the gas and the bulk of the optical matter. We could subsequently study the H$_\alpha$ morphological aspect of these isolated spiral galaxies.}
   {Using Fourier analysis, we focus on the modes of the spiral arms and also on the strength of the bars, computing the torques between the gas and newly formed stars (H$_\alpha$) and the bulk of the optical matter (Gunn r).}
   {We interpret the various bar/spiral morphologies observed in terms of the secular evolution experienced by galaxies in isolation. We also classify the different spatial distributions of star forming regions in barred galaxies. The observed frequency of particular patterns brings constraints on the lifetime of the various evolution phases. We propose an evolutive sequence accounting for the transitions between the different phases we could observe.
}
   {Isolated galaxies appear not to be preferentially barred or unbarred. Through numerical simulations, trying to fit the H$_\alpha$ distributions yields constraints on the star formation law, which is likely to differ from a genuine Schmidt law. In particular, it is probable that the relative velocity of the gas in the bar also needs to be taken into account.
}


   \maketitle
%

\section{Introduction} \label{sec:intro}
The presence of a bar in a spiral galaxy is a striking feature and, as such, is one of the fundamental elements of the first morphological classification done by Hubble. The bar (including ovals) frequency determined in galaxies depends on the wavelength of the images but is always higher than 65\%. Some estimations based on NIR images, not affected by extinction and tracing mainly the old population, reveal that as many as 90\% of the galaxies have bars \citep{1998MNRAS.299..672S}.

Numerical simulations have established that bars in gas-rich spiral galaxies are short-lived structures \citep{2002A&A...392...83B}. At least two mechanisms have been proposed to weaken the bars. The first one is the building of a large central mass concentration, due to the gas inflow to the centre through the negative torques exerted on the gas by the bar \citep{2005MNRAS.363..496A,2005MNRAS.364L..18B}. These torques are proportional to the phase shift between the gas and the stellar bar. It is well known that the gas is concentrated on the leading side of the stellar bar \citep[e.g.,][]{1963AJ.....68..278D}. The gravitational torque of the bar makes the gas lose angular momentum, driving it towards the centre and creating a central mass concentration. The latter is able to perturb the elongated orbits supporting the stellar bar, deflecting the stars passing close to the centre, so weakening the bar \citep{1990ApJ...363..391P}. However, \citet{2005MNRAS.364L..18B} have shown, using fully self-consistent simulations, that with gas parameters typical for normal spirals, the mass concentration is not sufficient to fully dissolve the bar, as was also claimed by \citet{2004ApJ...604..614S}.

The second proposed mechanism to weaken a bar are the bar torques themselves. The bar is a negative angular momentum wave, while the gas exerts a positive torque on the bar, due to the balance between action and reaction forces. The angular momentum lost by the gas is gained by the bar, which dissolves progressively. For typical Sb--Sc galaxies, the bar is destroyed in about 2 Gyr \citep{2002A&A...392...83B}. The observation of a high bar frequency from $z \sim 0$ to $z \sim 0.7$ \citep{2002ApJS..143...73E,2004ApJ...615L.105J,2005ApJ...632..217S} cannot thus be interpreted to support the existence of robust, long-lived bars. Instead, this supports the frequent renewal of bars. \citet{2004MNRAS.347..220B} have shown that interactions can only form bars in gas-poor galaxies, which is not the case for most spiral galaxies.

Bar renewal can occur when the disks of spiral galaxies are replenished in cold gas through external accretion, able to increase significantly the disk to bulge ratio. \citet{2002A&A...394L..35B} suggest that external accretion of gas in the disk of spirals plays a fundamental role in explaining the high fraction of barred spirals and the observed torque distribution. For instance, a recent study of M\,33 by \citet{2004A&A...425L..37B} illustrates that gas is feeding the galaxy through external accretion. Isolated, non-accreting spiral galaxies are very unlikely. To better understand the frequency of bars and their origin, measurements of the gravitational torques and bar forces in field galaxies have been made \citep{2002A&A...394L..35B,2004ApJ...607..103L,2004MNRAS.355.1251L,2004AJ....127..279B}, but not for a sample of isolated galaxies. Such a sample is provided by the AMIGA project ({\bf A}nalysis of the interstellar {\bf M}edium of {\bf I}solated {\bf GA}laxies), which constitutes a statistically significant template in the study of star formation and galaxy evolution in denser environments. We used as a starting point the Catalogue of Isolated Galaxies (CIG) compiled by \citet{1973AISAO...8....3K} (see Sect.~\ref{sec:samp}) for which a careful reevaluation of degree of isolation is presented in \citet{2007A&A...2..2V,2007A&A...470..505V}. Previous AMIGA papers evaluate, refine and improve the sample in different ways including  revised positions \citep{2003A&A...411..391L}, optical characterisation \citep{2005A&A...436..443V}, morphological revision \citep{2006A&A...449..937S}, analysis of mid- and far-infrared properties using data from the IRAS survey \citep{2007A&A...462..507L} and a study of the neutral (CO and \ion{H}{i}) gas  \citep{2005A&A...442..455E,2006.PhD.Thesis.E}. We also plan to study radio-continuum emission to determine current star formation rate and possible nuclear activity. The data are being released and periodically updated at {\tt http://www.amiga.iaa.es}.

In this paper we present a study of a carefully selected subsample of AMIGA, which is described in detail in Sect.~\ref{sec:samp}. The observations and main reduction steps are summarised in Sect.~\ref{sec:obs}. Then, we analyse, through Fourier analysis of gravitational potential and density, the intensity of the perturbations in each of the 45 galaxies (Sect.~\ref{sec:imAn}). Numerical simulations (Sect.~\ref{sec:numSimu}) of stars and gas in isolated galaxies give us the theoretically expected star formation. Our results are discussed in Sect.~\ref{sec:disc} and the conclusions are summarised in Sect.~\ref{sec:concl}.

\section{Sample of isolated spiral galaxies} \label{sec:samp}

As explained above, the AMIGA catalogue is based on the Catalogue of Isolated Galaxies \citep[CIG,][]{1973AISAO...8....3K}. This catalogue is a compilation of 1051 objects with apparent $B$ magnitude brighter than 15.7 and declination~$>$~$-3^{\circ}$. \citet{1973AISAO...8....3K} visually inspected the Palomar Sky Survey prints, trying to identify those galaxies in the Catalogue of Galaxies and Clusters of Galaxies \citep[CGCG,][]{1968cgcg.book.....Z} which have no near neighbours. An extensive study of the isolation of these galaxies is presented in \citet{2007A&A...2..2V,2007A&A...470..505V}, as a part of the AMIGA project, and discussed in Sect.~\ref{sec:disc}.

One of the aspects of our project was to study the H$_\alpha$ emission of isolated spirals in its different aspects. With this aim, and in order to avoid well known biases due to flux- or magnitude-limited samples and to end with a complete and homogeneous sample, we kept all the galaxies in a volume-limited sample, i.e. with observed recession heliocentric velocities $V_{\rm R}$:

\begin{center}
1500 km\,s$^{-1}$ $\le$ $V_{\rm R}$ $\le$ 5000 km\,s$^{-1}$.
\end{center}

It represents about one fourth (251 galaxies) of the whole CIG. Among them, 27 were early-type galaxies according to our new morphological classification \citep{2006A&A...449..937S}, hence rejected from the sample. We finally ended up with 224 galaxies with morphological types S0/a or later. While being a rather big sample allowing statistical studies of the properties of isolated spiral galaxies in the local Universe, its size still allows to perform H$_\alpha$ imaging for all the galaxies. These 224 galaxies constitute the main subsample for AMIGA star formation studies.

For the here presented study two further requirements had to be imposed:
\begin{enumerate}
\item Major axis $a \ge 1'$ to have a sufficient spatial resolution;
\item Inclination $i \le 50^\circ$ in order to obtain a sufficiently accurate deprojection.
\end{enumerate}

From our subsample, 48 galaxies follow these two criteria, of which 45 were observed and which constitute the subject of study in this paper. Table~\ref{tab:45c} reports information about the main characteristics of the selected galaxies, along with some technical details. The position angle ($PA$), inclination $i$ and rotation direction (assuming trailing spiral arms, the rotation will be set as positive if counter-clockwise) listed are those used to calculate the torques (24 counter-clockwise galaxies and 19 clockwise).

\section{Observations and data reduction} \label{sec:obs}

The 224 galaxies comprising our sample for star formation studies, including the 45 studied in this paper, have been imaged in H$_\alpha$ and Gunn r filters, in order to trace respectively the regions ionised by newly born stars (\ion{H}{ii} regions) and the older stellar component. The observations and data reduction will be described in more detail in a separate paper (Bergond et al., in prep.), and are just summarised here. For each galaxy, we selected the more appropriate red-shifted H$_\alpha$ filter (narrow-band filter, typically 50 \AA), based on its observed recession velocity and we used the Gunn r filter (broad-band filter, typically 870 \AA) to subtract the continuum contribution affecting the narrow-band filter. The Gunn r broad-band images trace the stellar component of the galaxy, and the H$_\alpha$ images show the young stars born from the gas (\ion{H}{ii} regions). The individual exposure times were 300 seconds for the Gunn r filter and 1200 seconds in H$_\alpha$. We applied small shifts between two successive exposures to be able to treat bad pixels on the chip. We applied the usual corrections (bias subtraction and division by twilight flat fields) to remove the instrumental signature.

The removal of the continuum contribution to the flux in the images taken through the narrow-band H$_\alpha$ filter is the most delicate task of the reduction process. \citet{2004A&A...414...23J} tested various methods: numerical integration to find the ratio of the filter profile integrals, photometry of standard spectrophotometric stars through the pair of filters to find the scaling factor, use of foreground stars in the narrow-band and continuum images of each galaxy. The two last methods gave the most consistent and accurate results. The latter method has two advantages: it takes into account any changes in the sky transparency between the two images and the possibility to use several stars improves the statistics and leads to the most accurate removal. Hence, we applied this last method to find the scaling factor to be applied to the images used for continuum subtraction. As the H$_\alpha$ wavelength is comprised in the Gunn r filter, a significant H$_\alpha$ emission would perturb the continuum removal. The uncertainties involved can be estimated in various ways. In a first approximation, with the hypothesis of negligible line emission (assuming a flat spectrum), the ratio of the width of the filters shows that the uncertainties amounted to about 6\% (using typical width of 50 and 870~\AA). Photometry of the total emission on the Gunn r and H$_\alpha$-continuum images confirms that the H$_\alpha$ emission generally represents a small fraction (typically 3\%) of the Gunn r emission. This latter value agrees with the estimations reported by \citet{2004A&A...414...23J} and \citet{2004A&A...426.1135K}.


\section{Image analysis} \label{sec:imAn}

In order to calculate the gravitational potential, density and torques in the plane of the disks of the galaxies, we processed the H$_\alpha$ and Gunn r corrected images as follows: the r band image (not scaled to H$_\alpha$) and the H$_\alpha$ continuum subtracted image were used. The Gunn r images were used in order to estimate the gravitational potential of the galaxies. This step is generally done with near-infrared images which trace more closely the gravitational potential (absence of extinction); unfortunately we do not possess these images for all the galaxies in our sample. Nevertheless, CIG\,1004 (NGC\,7479) is in the OSUBGS sample \citep{2002ApJS..143...73E} and its near-infrared potential had been estimated by \citet{2002A&A...394L..35B}. The potentials obtained from the Gunn r and near-infrared images are very similar (only the potential due to the optical spiral arm shows a slight enhancement in the Gunn r with respect to the near-infrared estimation) and hence the Gunn r image is a rather good approximation in order to derive the gravitational potential of a galaxy, in the absence of near-infrared information. We defined the centre as the maximum in luminosity near the geometrical centre. The centre is the same in the r and H$_\alpha$ image to avoid artificial torques between the two components. Then, we cut 512 $\times$ 512 pixels$^2$ subsections, as Fast Fourier Transform (FFT) will be applied on these images. Foreground galactic stars were removed in order to avoid a contamination in the density and potential derived from the luminosity of the pixels. The galaxy was deprojected to obtain the gravitational potential in the disk plane. The deprojection assumes that the outer disk of galaxies are circular and that the bulges are as flat as the disks. For galaxies inclined by less than 70\degr, \citet{2002A&A...394L..35B} found that the observational uncertainties on the inclination angles of the galaxies affect barely the estimation of the torques. Our conservative limit includes galaxies inclined less than 50\degr\ and assures us that the uncertainties in the deprojection do not affect our results.

Then, to determine the bar/arm force at each radius, we made use of the method developed by \citet{1993A&A...274..148G,2005A&A...441.1011G} to compute the gravitational potential, applying it to our red images, supposing a constant stellar mass-to-light ($M/L$) ratio. A Fourier-component analysis of the potential was performed, using the $m = 2$ component to get the tangential forces \citep{2002A&A...394L..35B} and the $m = 0$ component to get the axisymmetric forces. Their quotient gives us a measurement of the bar/arm force.

We estimated the average torque depending on the radius: using the gravitational forces and the young stars born from the gas (H$_\alpha$), we obtained the phase shift between gaseous arms and the potential well which creates the torques. Generally, due to the bar gravity torques, the gas flows into the centre (negative torque) from the corotation radius to the inner Linblad resonance (ILR) and flows outwards (positive torque) when outside the corotation radius, until the outer Lindblad resonance (OLR). This can be effectively observed for the galaxies in our sample. Finally, we estimated the angular momentum transfer and therefore the evolution time for bars in isolated galaxies (see subsection~\ref{sec:inter}).\\

The gravitational potential was decomposed as:
\begin{displaymath}
\Phi(r,\theta) = \Phi_0(r) + \sum_m \Phi_m(r) \cos (m \theta - \phi_m)
\end{displaymath}
where the strength of the $m$-Fourier component is $Q_m(r) = m \Phi_m / r | F_0(r) |$, and its global strength over the disk: $\max_r \left( Q_m(r) \right)$ \citep[e.g.,][]{1981A&A....96..164C}.\\

The (disk) surface density was decomposed as:
\begin{displaymath}
\mu(r,\phi) = \mu_0(r) + \sum_m a_m(r) \cos (m \phi - \phi_m (r))
\end{displaymath}
where the normalised strength of the Fourier component $m$ is $A_m(r) =  a_m / \mu_0 (r)$. The density Fourier analysis is complementary to that of the potential, which is less noisy but more global, and depending on the assumed 3D distribution of the mass.

The maximal torque at a given radius was defined by:
\begin{displaymath}
Q_{\rm T}(R) = {{F_{\rm T}^{\max}(R)} \over {F_0(R)}} = {{1\over R} \bigl{(} 
{{{{\rm{\partial}}\Phi(R,\theta)}}\over{{\rm{\partial}}\theta}}
             \bigr{)}_{\max} \over {{\rm{d}}\Phi_0(R)\over {\rm{d}}R}} 
\end{displaymath}
where $F_{\rm T}^{\max}(R)$ represents the maximum amplitude of the tangential force at radius $R$, and $F_{0}(R)$ is the mean axisymmetric radial force inferred from the $m=0$ component of the gravitational potential.\\

As an example, Figs.~\ref{fig:QATPaper0030} \&~\ref{fig:QATPaper1004} display the images and graphics of the potential, density and torques corresponding to CIG~30 \& CIG~1004, respectively. The first row shows the Gunn~r broadband filter (left) and H$_\alpha$ narrow-band filter (right) deprojected images, from which the potential, density and torques have been estimated. The second row corresponds to the amplitude (left) and the phase (right) of the Fourier components of the potential. The third row corresponds to the amplitude (left) and phase (right) of the Fourier components of the density. For the graphics of the potential and the density, the legend is the following: dashed line represents $m = 1$; full line represents the $m = 2$ component (and the sum also for the amplitude of the potential as it is always above and cannot be confused with the others); dot-dash-dot-dash represents $m = 3$; dotted line represents $m = 4$. The torques as a function of the radius of the galaxy is presented in the bottom left panel. The corotation occurs at the radius defined by the change of sign for dL/L. And dL/L shows that in one rotation, the gas loses a non negligible fraction of its angular momentum, which can bring constraints on the life time of the morphological patterns of the galaxy. The legend in the bottom right corner gathers the CIG number, the observed recession velocity (in km~s$^{-1}$), the optical diameter $D_{25}$ (in $'$), the blue luminosity (logarithm, in $L_\odot$; see \citet{2007A&A...462..507L}, their Table~3), the morphological type (from NED) and the group to which the galaxy belongs (see Sect.~\ref{sec:disc}). The Gunn~r and H$_\alpha$ images, along with the graphics of the potential, density and torques of the full sample of the 45 galaxies are 
also available at {\tt http://www.amiga.iaa.es/publications}.

\newcommand{\PageImagePaper}[8]{
\newpage

\begin{figure*}
\hfill
\begin{minipage}[t]{0.42\textwidth}
\includegraphics[width=\textwidth]{7650-#1a_resamp.eps}
\end{minipage}
\hfill
\begin{minipage}[t]{0.42\textwidth}
\includegraphics[width=\textwidth]{7650-#1b_resamp.eps}
\end{minipage}
\hfill
\vfill
\begin{minipage}[t]{0.40\textwidth}
\includegraphics[width=\textwidth]{7650-#1c.eps}
\end{minipage}
\hfill
\begin{minipage}[t]{0.40\textwidth}
\includegraphics[width=\textwidth]{7650-#1d.eps}
\end{minipage}
\hfill
\vfill
\begin{minipage}[t]{0.40\textwidth}
\includegraphics[width=\textwidth]{7650-#1e.eps}
\end{minipage}
\hfill
\begin{minipage}[t]{0.40\textwidth}
\includegraphics[width=\textwidth]{7650-#1f.eps}
\end{minipage}
\hfill
\vfill
\begin{minipage}[c]{0.38\textwidth}
\includegraphics[width=\textwidth]{7650-#1g.eps}
\end{minipage}
\hfill
\begin{minipage}[c]{0.12\textwidth}
\end{minipage}
\hfill
\begin{minipage}[c]{0.42\textwidth}
\caption[CIG #2.]{{\bf #3 (CIG #2).}\newline
V = #4 km\,s$^{-1}$\newline
Optical diameter: #5 arcmin.\newline
Blue luminosity: #6\newline
Morphology: #7\newline
Group: #8\newline
{\small For the potential, $Q_{\rm b} = Q_m$, the normalised amplitudes of the $m$ Fourier modes. For the graphics of the potential and the density, the dashed line represents $m = 1$; full line represents the $m = 2$ component (and the sum also for the amplitude of the potential as it is always above and cannot be confused with the others); dot-dash-dot-dash represents $m = 3$; dotted line represents $m = 4$.}
}\label{fig:QATPaper#2}
\end{minipage}
\hfill
\vfill
\end{figure*}
}
\PageImagePaper{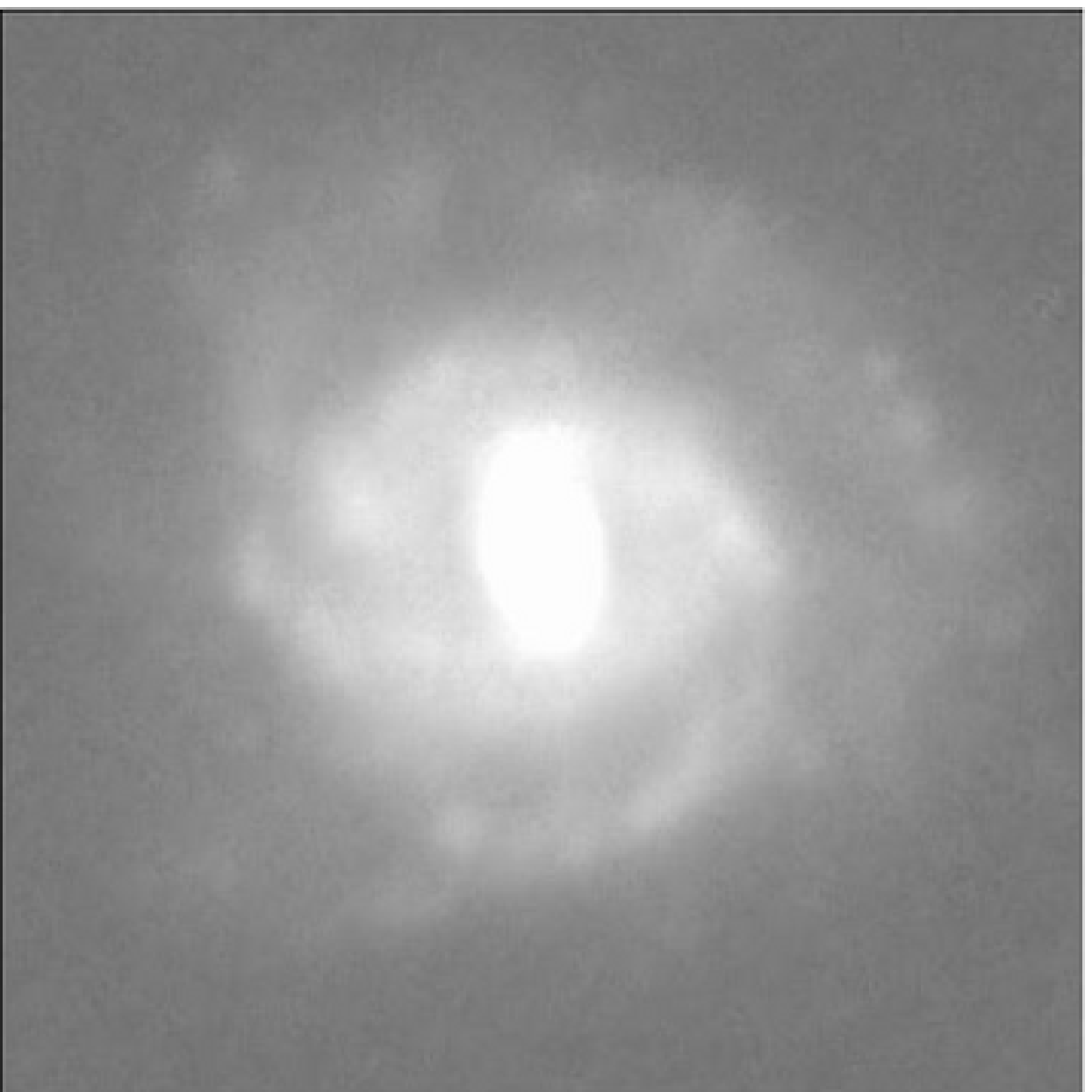}{0030}{IC\,35}    {4586}{1.0}{9.82}{Scd:}{E}
\PageImagePaper{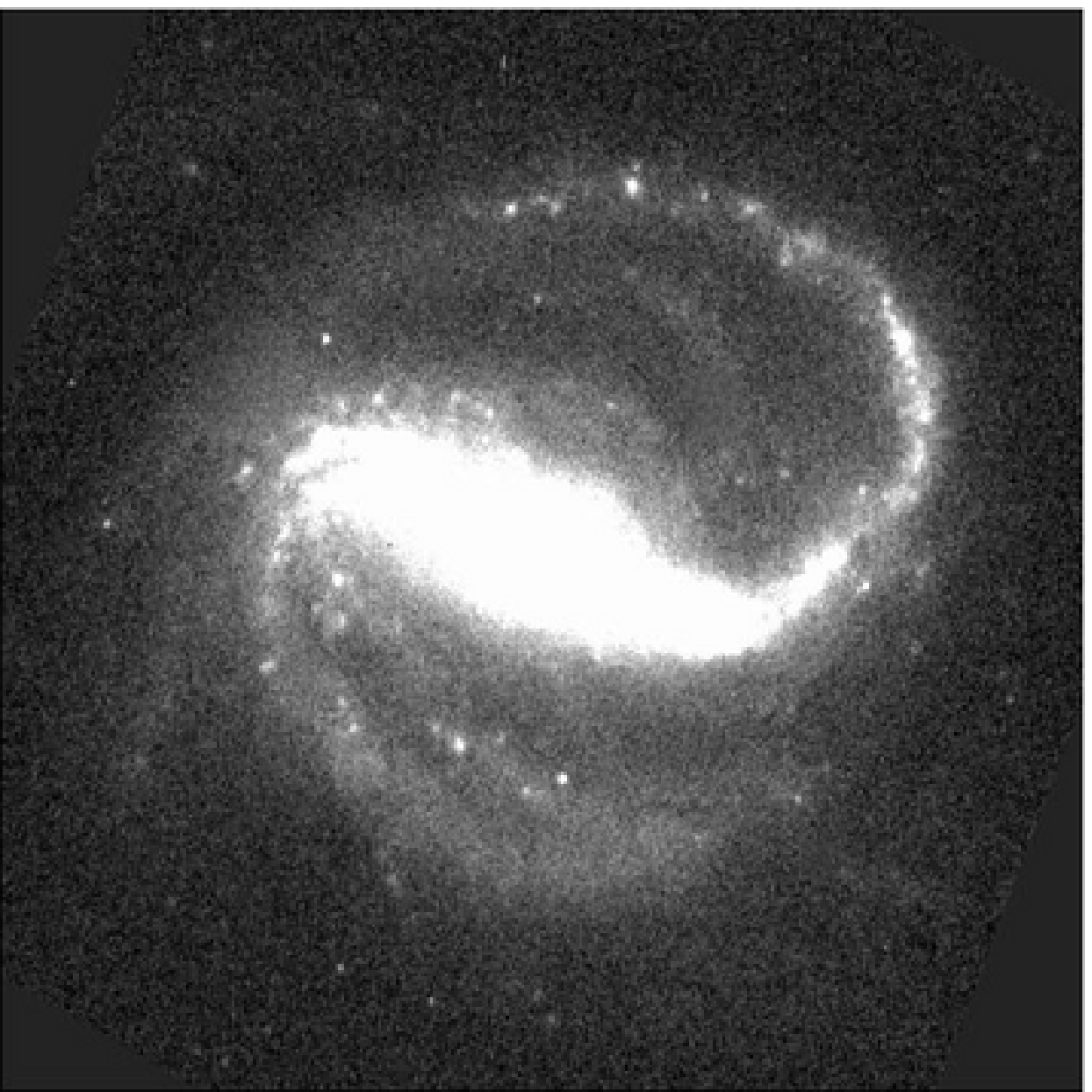}{1004}{NGC\,7479} {2376}{4.1}{10.40}{SB(s)c; LINER Sy2}{G}

\section{Numerical simulations} \label{sec:numSimu}

\subsection{Methods}

In order to understand the observed H$_\alpha$ distributions, and the different phases identified, we performed $N$-body simulations with stars and gas, including star formation. Since we want to explore many physical parameters, we chose to carry out 2D simulations, which should capture the essential of the bar evolution, and location of star formation in these isolated galaxies. The 3D components, bulge and dark matter halo, are therefore considered as rigid and spherical potentials, in which the disk component evolves. The rigid or live character of the spheroidal components would make a significant effect, by exchanging angular momentum, only if their mass is relatively significant. But in the inner parts of galaxies, the dark matter halo is not dominant, and thus this phenomenon would only induce small changes. The same is true for the spheroidal bulges, which are relatively small in mas, in these mostly late-type galaxies. Self-gravity is only included for the disk (gas + stars). 2D $N$-body simulations were carried out using the FFT algorithm to solve the Poisson equation, with a Cartesian grid, varying from $256 \times 256$ to $512 \times 512$ (useful grid, free of periodic images). Two spatial resolutions were selected to appreciate its influence on the star formation physics. The cell size is then from 62.5 to 125~pc, and the total size of the grid is 32~kpc. The softening length of the gravity has the characteristic scale of the cell (62.5 to 125~pc). More details on the numerical techniques can be found in \citet{1990A&A...233...82C}.

The stellar component is represented by 100\,000 or 400\,000 particles, and the gas component by 40\,000 and 160\,000 for the low and high resolutions adopted, respectively.

The bulge is modelled as a rigid spherical potential with Plummer shape:
$$
\Phi_{\rm b}(r) = - { {G M_{\rm b}}\over {\sqrt{r^2 +r_{\rm b}^2}} }
$$
$M_{\rm b}$ and $r_{\rm b}$ are the
mass and characteristic radius of the bulge.

\begin{figure}
\includegraphics[angle=-90,width=\columnwidth]{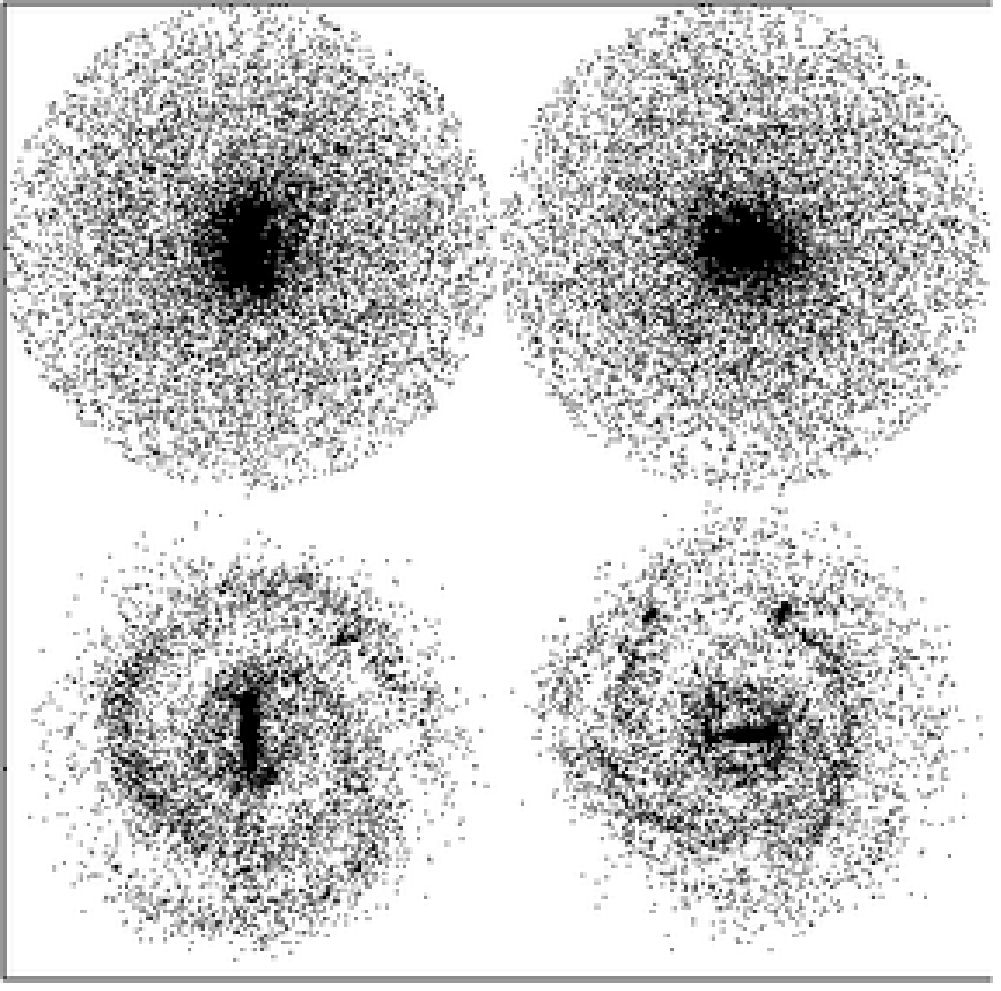}
\caption{Particle plots of the stars (top) and gas (bottom) at the epoch
$T=1.1$ Gyr (left) and $T=1.5$ Gyr (right), in the simulation of Run 1. The scale of the square is 64 kpc.} \label{fig:run1-sg1}
\end{figure}

\begin{figure}
\includegraphics[angle=-90,width=\columnwidth]{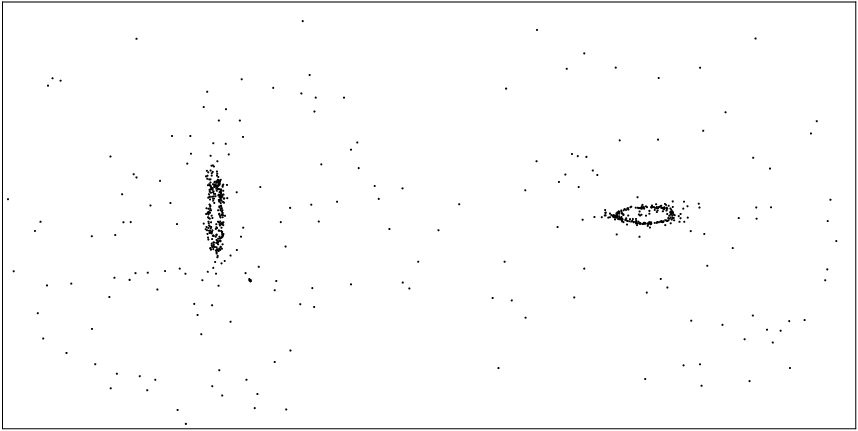} 
\caption{Plots of recently formed stars (less than 200 Myr ago), at the epoch $T=1.1$ Gyr (left) and $T=1.5$ Gyr (right), in the simulation of Run 1. The scale of the square is 32 kpc. These plots should give an idea of the H$_\alpha$ map.} \label{fig:run1-ha2}
\end{figure}

\begin{figure}
\includegraphics[angle=-90,width=\columnwidth]{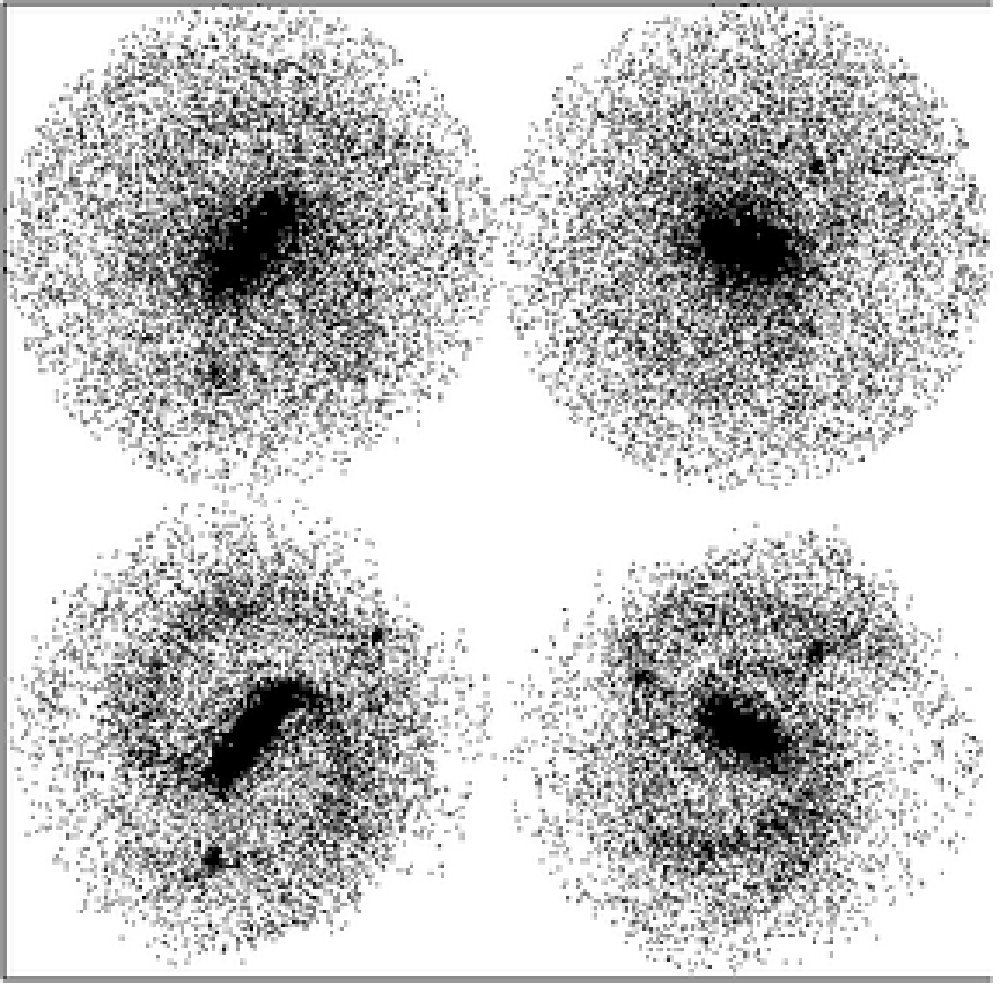}
\caption{Particle plots of the stars (top) and gas (bottom) at the epoch $T=0.6$ Gyr (left) and $T=1.0$ Gyr (right), in the simulation of Run 4. The scale of the square is 64~kpc.} \label{fig:run4-sg1}
\end{figure}

The stellar disk is initially a Kuzmin-Toomre disk of surface density:
$$
\Sigma(r) = \Sigma_0 ( 1 +r^2/r_{\rm d}^2 )^{-3/2}
$$
truncated at 15~kpc, with a mass  $M_{\rm d}$. It is initially quite cold, with a Toomre $Q_{\rm T}$ parameter of 1. The halo is also a Plummer sphere, with mass $M_{\rm h}$ and characteristic radius $r_{\rm h}$. The time steps are 0.5 and 1 Myr. The initial conditions  of the runs described here are given in Table~\ref{tab:ci}. The particle plots of the stars and gas as well as the recently formed stars are depicted in Figs.~\ref{fig:run1-sg1}~\&~\ref{fig:run1-ha2} and Figs.~\ref{fig:run4-sg1}~\&~\ref{fig:run4-ha2}, for Run~1 and Run~4, respectively.

\begin{table}
\caption{Initial conditions for the numerical simulations. $N_{\rm p}$ is the number of particles considered, $r_{\rm {b/d/h}}$ and $M_{\rm {b/d/h}}$ the galaxy bulge/disk/halo radii and masses,
and $f_{\rm {el}}$ the collision elasticity factor.} \label{tab:ci}
\begin{flushleft}
\begin{tabular}{ccccccccc} \hline \hline
Run  &$N_{\rm p}$     & $r_{\rm b}$& $M_{\rm b}$    & $r_{\rm d}$ & 
$M_{\rm d}$   & $r_{\rm h}$ & $M_{\rm h}$  & $f_{\rm {el}}$  \\
id.        & $\times$$10^3$ & kpc  & $^{(1)}$   & kpc&  $^{(1)}$  & kpc& $^{(1)}$       &        \\
\hline
Run 0  & 100   &  1.1 &  2.5 &  4.4 & 8.0 & 16.0  & 7.2       & 0.65    \\
\hline
Run 1  & 100   &  1.1 &  6.8 &  4.4 & 8.0 & 16.0  & 11.8       & 0.65    \\
Run 2   & 100  &  1.1 &  6.8 &  4.4 & 8.0 & 16.0  & 11.8        & 0.85    \\
\hline
Run 3  & 400   &  1.1 &  6.8 &  4.4 & 8.0 & 16.0  & 11.8       & 0.65    \\
Run 4   & 400  &  1.1 &  6.8 &  4.4 & 8.0 & 16.0  & 11.8        & 0.85    \\
\hline
\end{tabular}
{\scriptsize{$^{(1)}$Masses are in  10$^{10} M_{\sun}$.}}
\end{flushleft}
\end{table}

The gas is treated as a self-gravitating component in the $N$-body simulation, and its dissipation is treated by a sticky particle code, as in \citet{1985A&A...150..327C}. The initial gas-to-total mass ratio ($F_{\rm{gas}}$) in the disk ranges between 6 and 14\%, since the star formation in the simulation is capable of reducing $F_{\rm gas}$ to a final lower value. The mass of one gas particle therefore varies between $8\times10^4$ and $3\times10^5 M_{\sun}$.

The initial distribution of gas in the model is an exponential disk, truncated at 15~kpc, and with a characteristic radial scale of 6~kpc. Initially, its velocity dispersion corresponds to a Toomre $Q_{\rm T}$ parameter of 1. The gas clouds are subject to inelastic collisions, with a collision cell size between 60 and 120~pc (region where particles are selected to possibly collide). This corresponds to a lower limit for the average mean free path of clouds between two collisions. The collisions are considered every 5 to 10 Myr. In a collision, the sign of the relative cloud velocities is reversed and the absolute values are reduced: relative velocities after the collision are only $f_{\rm {el}}$ times their original value, the elasticity factor $f_{\rm{el}}$ being between 0.65 and 0.85, as indicated in Table~\ref{tab:ci}. The dissipation rate is controlled by this factor. All gas particles have the same mass.

Star formation is taken into account following a generalised Schmidt law: the star formation rate is proportional to the volume density to the power $n=1.2$, provided that the gas volumic density is larger than $\rho_{\rm{gas}}=1$ H-atom cm$^{-3}$, i.e. the rate of gas mass transformed into stars is d$m \propto \rho_{\rm{gas}}^{1.2} {\rm d}t$. To compute this rate, at regular intervals of d$t= 5$--10 Myr, the gas density is averaged in each cell, and the probability of the gas particles being transformed into stars is computed by:
$$
P = {\rm d}m/M_{\rm{cell}}
$$
for all particles in this cell, of mass $M_{\rm{cell}}$. Each new star formed has exactly the same mass as each gas particle, about 3 times smaller than any old stellar particle. This simple scheme corresponds to an instantaneous recycling of matter, since the continuous mass-loss from recently formed stars is not followed.  The rate of star formation is normalised so that in unperturbed runs (without galaxy interaction, galaxies are quiescently and regularly forming stars), the timescale for consumption of half of the gas mass is of the order of 2 Gyr (SFR $\sim$ 1--2  $M_{\sun}$/yr). At each star formation event, the neighbouring gas particles are given a small extra velocity dispersion of the order of $\sim$ 10 km\,s$^{-1}$.

To reproduce a model equivalent of the H$_\alpha$ maps, we plot only the recently born stars, still able to ionise their surroundings. In order to have more statistics, we choose to plot all new stars of age less than 200~Myr (Figs. \ref{fig:run1-ha2} \& \ref{fig:run4-ha2}).

\subsection{Results}

We tried several runs corresponding to the various morphological types of the Hubble sequence, varying the bulge to disk mass ratio, the halo to disk mass ratio and the gas content. This had the consequence to vary the dynamical stability of the disk: the bar appears within different time-scales, and with mainly two different morphologies for, respectively, early-types and late-types. In the case of late-types, the bar develops without any resonant ring, and quickly the gas infalls to the center, and destroys the bar. The H$_\alpha$ morphology has then a bar shape during a short period, and quickly peaks at the center. For early-types, the high central mass concentration allows the formation of inner Linblad resonances, and gaseous resonant rings are developped inside the bar. Also the life-time of the bar is long enough to develop a resonant ring at the 4:1 resonance near corotation, surrounding the bar (Fig.~\ref{fig:run1-sg1}). However the gas density there is not enough to produce substantial star formation (Fig.~\ref{fig:run1-ha2}).

\begin{figure}
\includegraphics[angle=-90,width=\columnwidth]{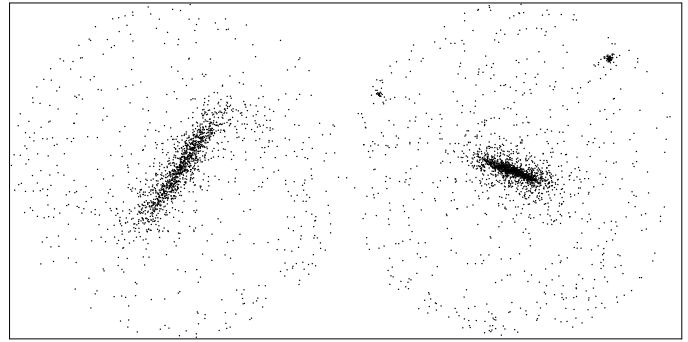} 
\caption{Plots of recently formed stars (less than 200 Myr ago), at the epoch $T=0.6$ Gyr (left) and $T=1.0$ Gyr (right), in the simulation of Run 4. The scale of the square is 32 kpc. Note the orientation of the young star component, which keeps the phase shift of the gas component during some time, and is leading with respect to the old stellar bar of Fig.~\ref{fig:run4-sg1} by about 3 degrees.} \label{fig:run4-ha2}
\end{figure}

Whatever the morphological type of the galaxy, the models always show a first stage where the main star formation is aligned with the bar, and a second stage, where the star forming regions concentrate towards the nucleus.  Variations of the degree of dissipation of the gas component, through the variation of collision elasticity $f_{\rm{el}}$, does not change the picture, as shown by the comparison between Run 1 and Run 4 in Figs.~\ref{fig:run1-ha2} and \ref{fig:run4-ha2}. The corresponding H$_\alpha$ maps that can then be predicted are either of the group {\bf G}, as defined in the next section, or the group {\bf H}. But it is not possible to obtain the most frequently observed morphology of the group {\bf E}.  In other words, the equivalent H$_\alpha$ maps are predicted to correlate tightly to the gaseous maps. This is expected from the type of recipe adopted for the star formation rate, i.e. a Schmidt law, with a density threshold. We defer the discussion of these results to Sect.~\ref{sec:inter}.

\section{Discussion} \label{sec:disc}

Our sample being 94\% complete (we are only missing data for 3 galaxies over the 48 we had selected, which could not have been observed), we can describe statistically significant characteristics for a sample of isolated galaxies.

\subsection{Isolation of the galaxies}

\citet{2007A&A...2..2V,2007A&A...470..505V} lead an extensive study to revise the degree of isolation of 950 CIG galaxies. Within a projected radius of 0.5~Mpc, all the neighbours that could exert an influence on the evolution of the CIG galaxies were catalogued. Parameters quantifying the degree of isolation of each CIG galaxy were defined. A particularly relevant parameter, $Q$, measures the strength of the tidal forces affecting the central galaxy, normalised to the inner binding forces of the galaxy itself. In the subsample of 45 galaxies defined in the present study, three of the galaxies present tidal forces amounting to more than 10\% ($Q \ge -1$) of their inner binding forces. This means that their evolution could have been seriously affected by nearby companions. These three galaxies are CIGs~80, 85 and 1001.

Also, other eleven galaxies present values of the external forces amounting to more than 1\% of their internal binding forces ($Q \ge -2$), which is the limit from which the environment can partially influence the evolution of the central galaxy \citep{1984PhR...114..321A}. These galaxies are CIGs~59, 84, 96, 176, 376, 382, 512, 645, 808, 812 and 1004. Nevertheless, these galaxies are isolated from similar size neighbours \citep[within a factor 4 in size, as originally defined by][]{1973AISAO...8....3K}. A statistical rejection of the projected distant neighbours (with a difference in recession velocity of almost 20\,000~km~s$^{-1}$) reveals that all but one (CIG~176) can, indeed, be galaxies whose evolution is dominated by their internal properties. More data on the recession velocities of the neighbours are needed in order to access a definitive picture of the environment of each CIG galaxy.

\subsection{Percentage of bars}

To know the percentage of barred galaxies in a sample of isolated galaxies, we visually classified the 45 galaxies in three subsamples, as a function of the presence or absence of bar:
\begin{description}
\item{\bf Bar} {\sl 27 galaxies -- CIGs 30, 53, 66, 96, 116, 176, 188, 250, 267, 376, 382, 512, 645, 652, 660, 661, 712, 754, 808, 840, 854, 862, 931, 935, 1001, 1004, 1039.}
\item{\bf No bar} {\sl 15 galaxies -- CIGs 50, 59, 80, 84, 85, 217, 281, 291, 359, 700, 750, 812, 875, 924, 992.}
\item{\bf Intermediate} {\sl 3 galaxies -- CIGs 68, 575, 744.}
\end{description}

The barred galaxies represent 60\% of our sample (the bar can be seen either in Gunn r or in H$_\alpha$). As we are using optical images, this result is consistent with the studies presented in Sect.~\ref{sec:intro}. The unbarred galaxies represent 33\% of the sample. Three galaxies have an intermediate stage bar, similar to the ``SAB'' kind, they represent 7\% of the total. As the morphological recognition is highly dependent both on the material inspected and on the human factor, some of the galaxies that we classified as ``barred'' would have shifted into the ``intermediate'' category following the classification done by \citet{1963ApJS....8...31D}. Nevertheless, we can infer that the isolated galaxies span the whole range of bar morphologies, in quantities similar to the galaxies in denser environments. Isolated galaxies are not preferentially barred or unbarred galaxies. This result is marginally in contradiction with the study by \citet{2004A&A...420..873V}, who estimate that bars are twice more frequent in perturbed galaxies compared to isolated galaxies, especially for early-types. Also \citet{1990ApJ...364..415E} find more bars in a sample of binary galaxies, and also more early-types.

\subsection{Phase shift between gas and stellar components}

\begin{figure*}
\begin{minipage}[t]{0.33\textwidth}
\includegraphics[width=\textwidth]{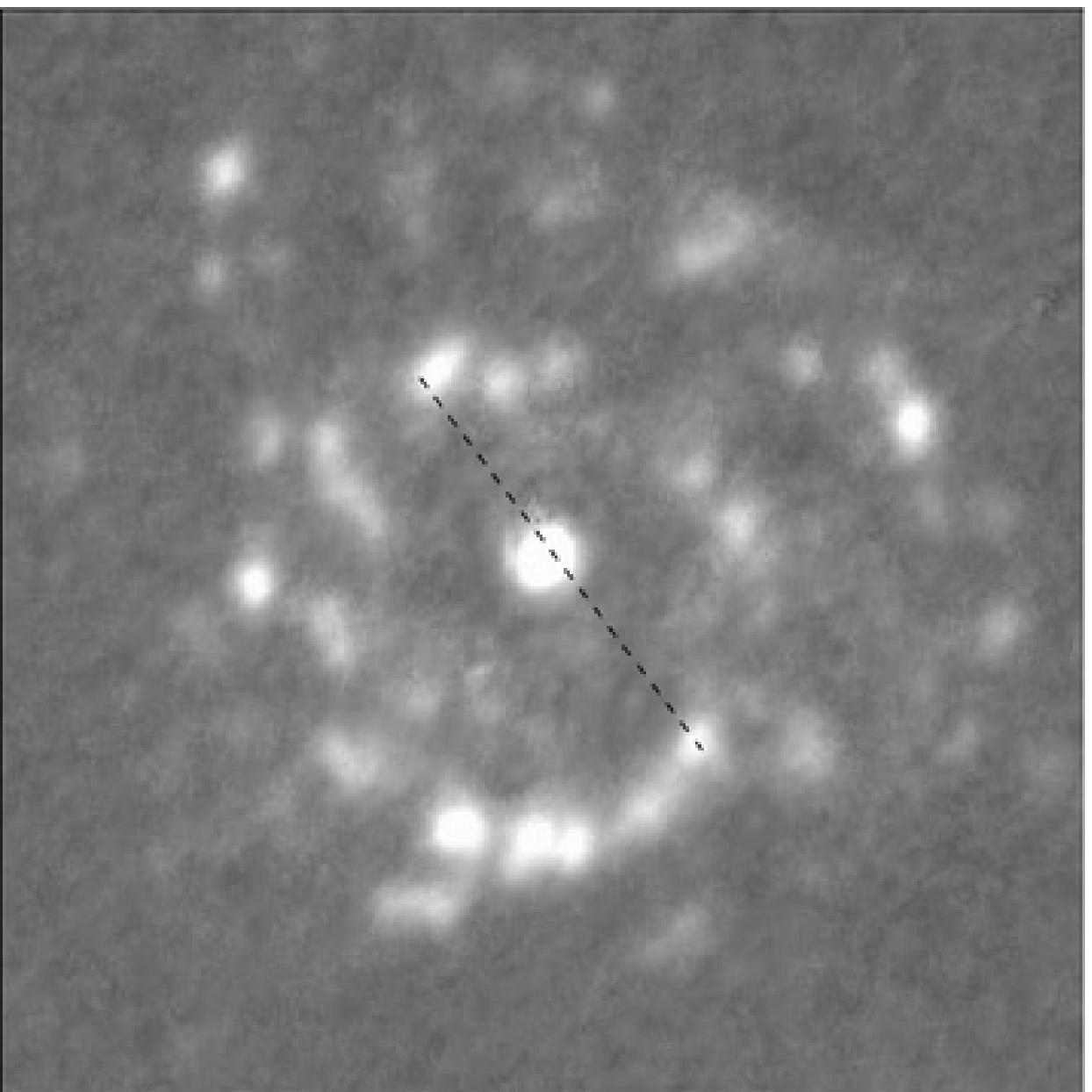}
\end{minipage}
\begin{minipage}[t]{0.33\textwidth}
\includegraphics[width=\textwidth]{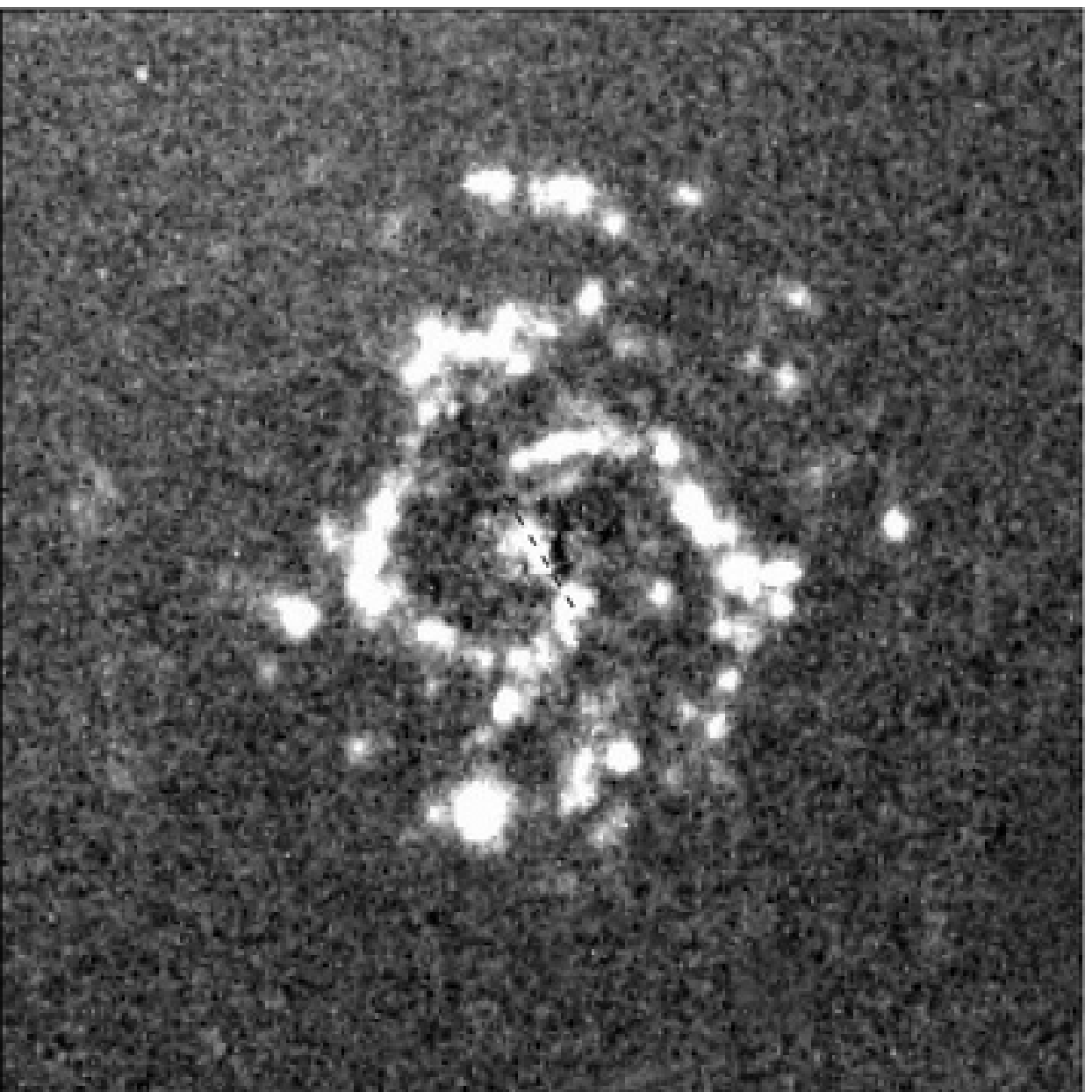}
\end{minipage}
\begin{minipage}[t]{0.33\textwidth}
\includegraphics[width=\textwidth]{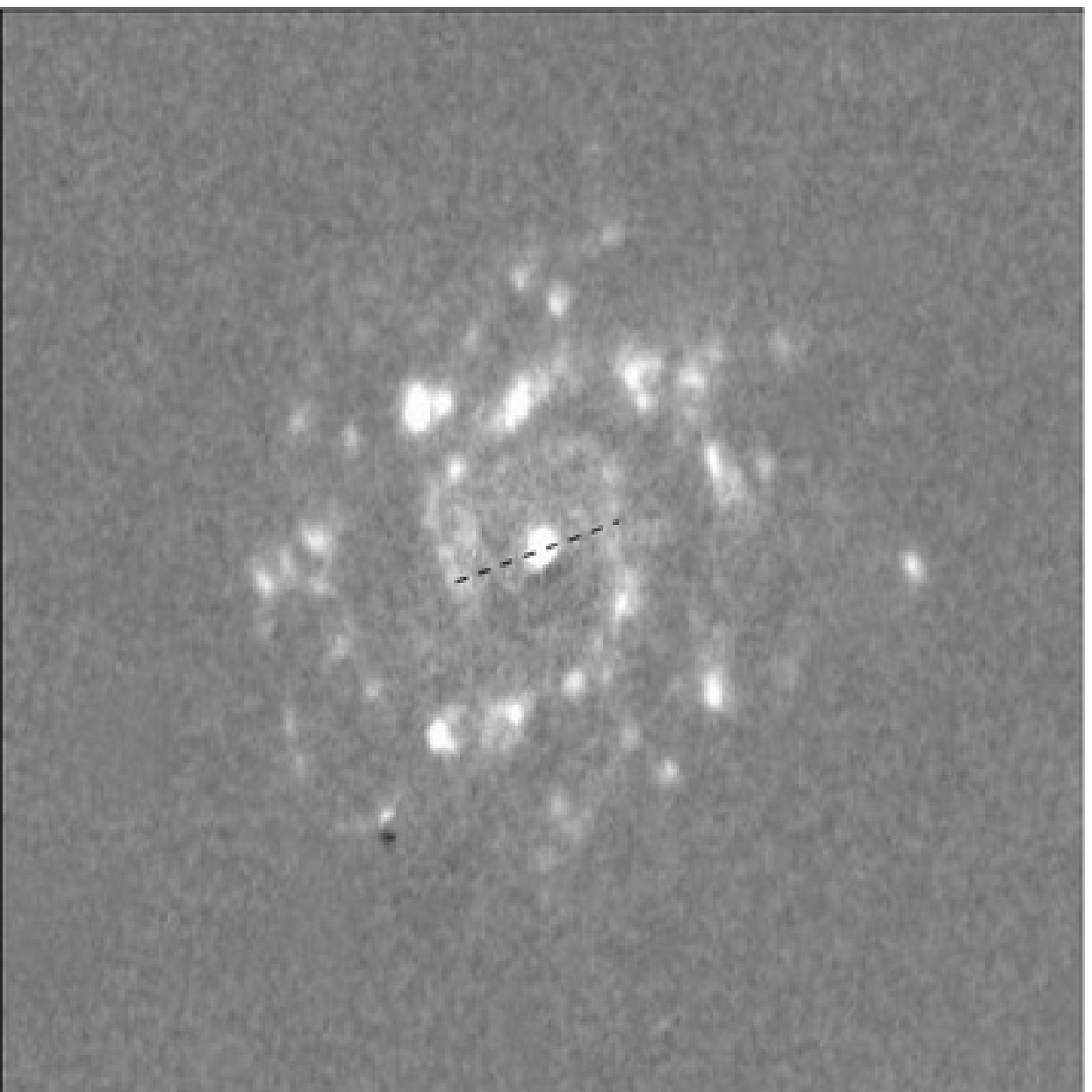}
\end{minipage}
\begin{minipage}[t]{0.33\textwidth}
\includegraphics[width=\textwidth]{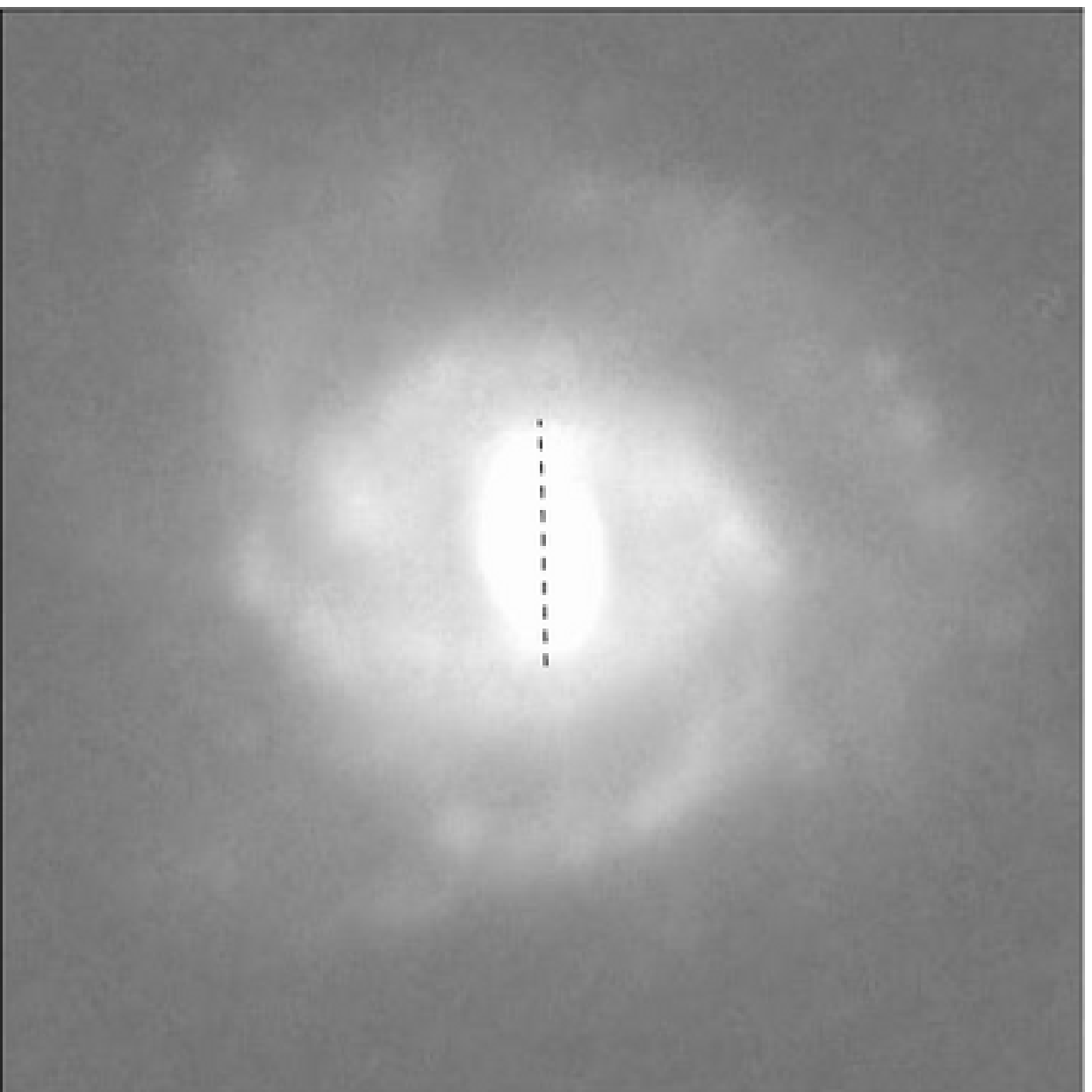}
\end{minipage}
\begin{minipage}[t]{0.33\textwidth}
\includegraphics[width=\textwidth]{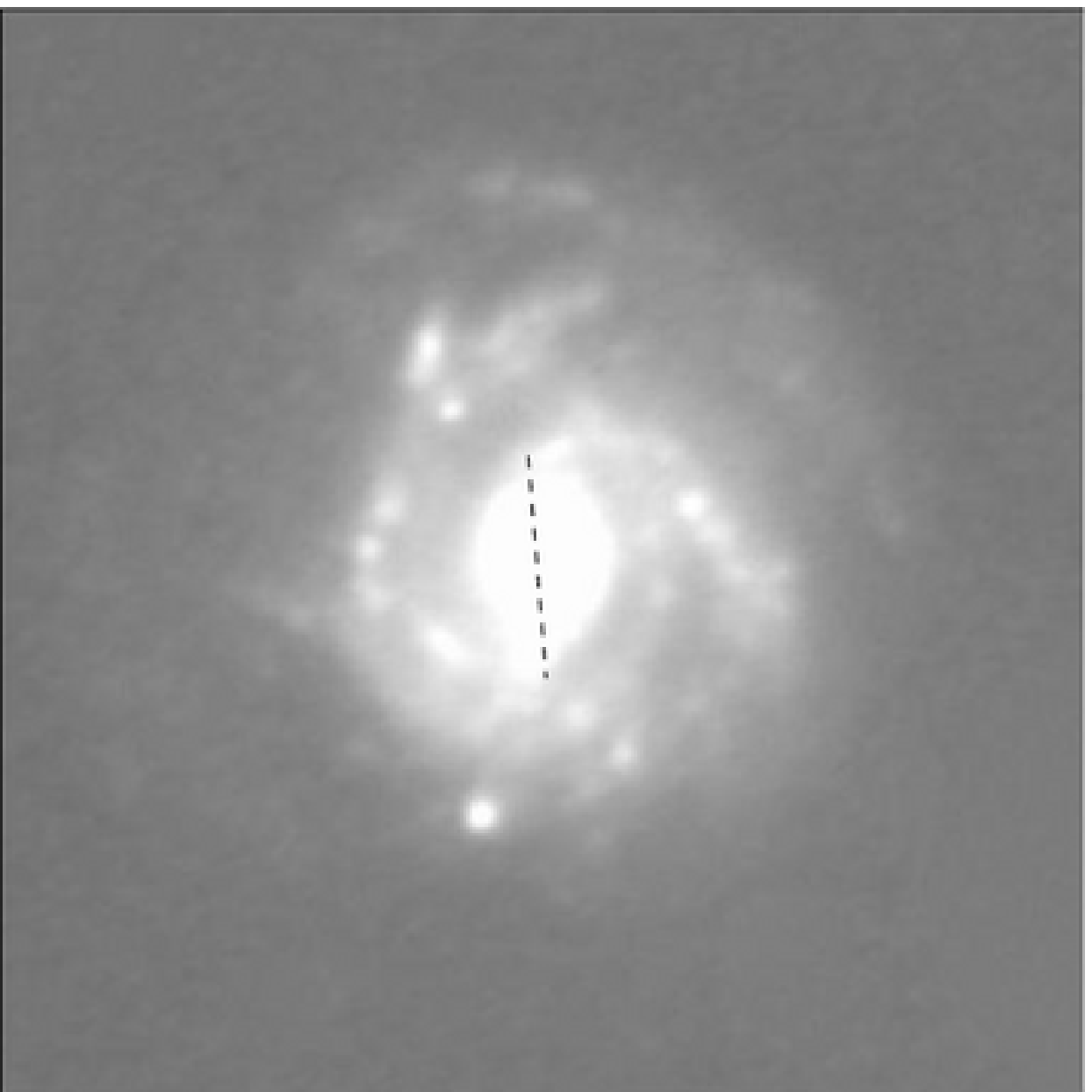}
\end{minipage}
\begin{minipage}[t]{0.33\textwidth}
\includegraphics[width=\textwidth]{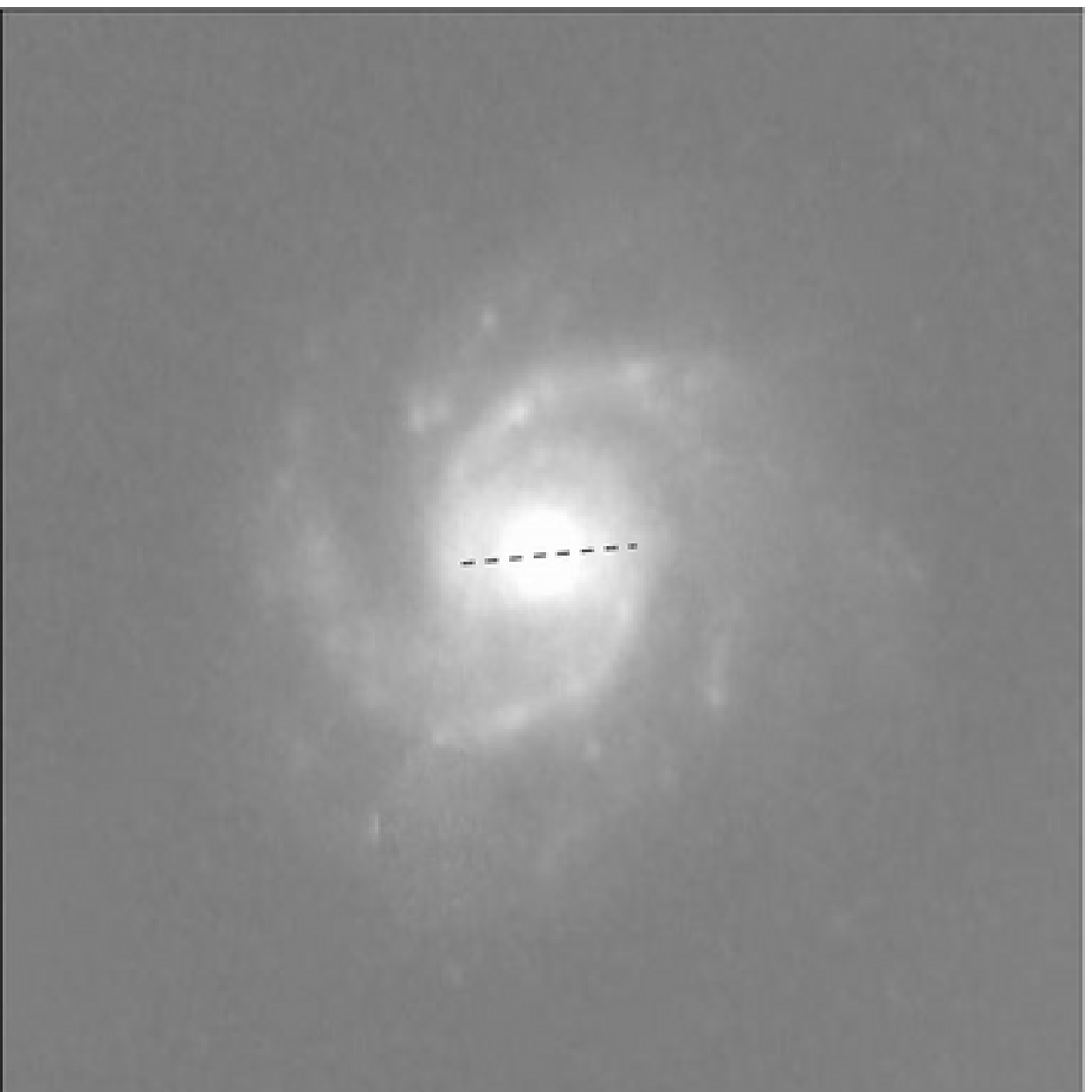}
\end{minipage}
\caption{Examples of shift angles between the H$_\alpha$ and stellar bars. The dash line represents the positon angle of the perturbation found in the image analysis. {\it Left:} Deprojected images of CIG~30 in H$_\alpha$-continuum (upper panel) and Gunn r (lower panel). {\it Middle:} Deprojected images of CIG~376 in H$_\alpha$-continuum (upper panel) and Gunn r (lower panel). {\it Right:} Deprojected images of CIG~840 in H$_\alpha$-continuum (upper panel) and Gunn r (lower panel).}\label{fig:shifts}
\end{figure*}

We now focus on the existence of shift angle between the location of the gas and the location of the older stellar component, which has already been noted by \citet{1997A&A...326..449M} in a sample of eleven galaxies \citep[for other examples, see also][]{1963AJ.....68..278D,1996ASPC...91...44P}. It was sometimes difficult to define where exactly the arms begin because the H$_\alpha$ could be very clumpy, but frequently, we saw bright H$_\alpha$ knots at the end of the bars, even when there was no emission in the bar. Consequently, we could use the star formation spots at the end of the bars, to define the starting point of the spiral arms. The H$_\alpha$ emission was always leading with respect to the bar in the Gunn r image. The most evident cases of bar shifts between the Gunn r and the H$_\alpha$ images are listed hereafter (in parentheses: an estimation of the shift angle in degrees): CIG~30 (30$^\circ$); CIG~53 (10$^\circ$); CIG~66 (5$^\circ$); CIG~96 (10$^\circ$); CIG~176 (10$^\circ$); CIG~376 (30$^\circ$); CIG~512 (10$^\circ$); CIG~840 (15$^\circ$); CIG~1004 (5$^\circ$). Three cases are illustrated in Fig.~\ref{fig:shifts}, where the orientations of the H$_\alpha$ and stellar bars have been marked by dashed black lines.

\subsection{Evolutive sequence} \label{sub:grp}

\subsubsection{Classification}

As some characteristics were frequently found among the galaxies of our sample, we chose to group the galaxies presenting similar features. In decreasing frequency order, we defined the three main groups:
\begin{description}
\item{Group \bf E} {\sl 19 galaxies -- CIGs 30, 53, 66, 80, 96, 176, 217, 281, 291, 376, 382, 660, 812, 840, 862, 931, 992, 1001, 1039.}\\*
The principal features of this group are the following: a strong central peak in the H$_\alpha$ emission; no H$_\alpha$ emission in the bar (for the barred galaxies in the Gunn r image); bright H$_\alpha$ knots at the end of the bar/beginning of the spiral arms; H$_\alpha$ emission along the spiral arms, generally clumpy.
\item{Group \bf F} {\sl 9 galaxies -- CIGs 50, 59, 84, 188, 267, 661, 808, 924, 935}.\\*
This group is constituted by galaxies with less gas, having a smoother morphology. These galaxies do not present any central emission spot in H$_\alpha$.
\item{Group \bf G} {\sl 8 galaxies -- CIGs 116, 250, 645, 700, 712, 754, 854, 1004.}\\*
This group gathers galaxies presenting H$_\alpha$ emission in the bar. CIG 700 presents a very faint emission, but a bar can be distinguished.\\
\end{description}

\noindent Five of the galaxies of our sample did not fit in any of the main groups defined above, while four others presented characteristics mixing features from two of the groups:
\begin{description}
\item{Group \bf H} {\sl 3 galaxies -- CIGs 68, 359, 875.}\\*
This group is mostly constituted by early types galaxies, with very few emission in H$_\alpha$ and when it occurs, only in the centre.
\item{Group \bf Irr} {\sl 2 galaxies -- CIGs 85, 575.}\\*
This group gathers very irregular galaxies which do not fit particularly in any of the previous groups defined above.
\item{Group \bf EG} {\sl 3 galaxies -- CIGs 512, 652, 744.}\\*
This group is a transition between the {\bf E} and {\bf G} groups presented above: the galaxies exhibit fragments of bars. The central emission of CIG 744 is very clumpy but seems to follow  the shape of a bar, this is why we included it in this group.
\item{Group \bf EF} {\sl 1 galaxy -- CIG 750.}\\*
CIG 750 does not have a clear central peak in H$_\alpha$ but presents all the other features of the galaxies in the group {\bf E}: clumpy H$_\alpha$ emission along the spiral arms. We call this group {\bf EF} as a result of the mixed features presented by CIG~750.\\
\end{description}

\subsubsection{Interpretation} \label{sec:inter}

Many $N$-body simulations of galaxies with gas have revealed that bars are not long-lived in late-type galaxies \citep[e.g.][]{1993A&A...268...65F,1998MNRAS.300...49B,2002A&A...392...83B}. \citet{2005MNRAS.364L..18B} have shown that the bar dissolution is due to the gas inflow towards the centre, and that bars can be renewed by external gas accretion. Our goal here is to get more insight in the various dynamical processes that could be involved in the interpretation of the different observed morphologies as different stages of an evolutive sequence. In our sample, we can identify various main steps occurring during the formation and evolution of the bars.

By gravitational instability, a galaxy accretes gas from the inter-galactic medium which makes it unstable for bar formation. The bar creates a torque which drives the gas inflow towards the centre. This phase corresponds to our identified {\bf G} phase (see, for instance, Fig.~\ref{fig:QATPaper1004} for a typical galaxy belonging to this group).

The second step is a transition between the {\bf G} and {\bf E} phases: the gas inflows towards the centre while a ring is slowly forming at the resonance (pseudo-ring due to the winding of the spiral arms, at the ILR).

In a third step, the gas is progressively depopulated from the bar, and accumulates first in the very centre of the galaxy (or a very small nuclear ring, at ILR), and also at the ultra harmonic resonance (UHR), near the corotation. The gas there is quite stable, with very small  relative velocity with respect to the bar pattern, and the star formation could be quite efficient in these regions. This corresponds to our identified more frequent phase, the {\bf E} phase, where galaxies could spend more than 40\% of their life-span (see, for instance, Fig.~\ref{fig:QATPaper0030} for a typical galaxy belonging to this group).

Since the gas infall destroys the bar, the latter becomes progressively weaker and weaker: the {\bf F} phase is reached. The stars in the centre become an old population, contributing to increase the bulge mass. Without more gas fuelling, the H$_\alpha$ spot in the centre is fading away in 10$^8$ years (OB stars). The frequency of the {\bf F} phase means that a galaxy spends typically 20\% of its life in this stage, where the bar is weakened or destroyed.

The various classes proposed in the previous section were based on the H$_\alpha$ map morphologies. By plotting images of the recent stars formed in the simulations, we have tried to predict those H$_\alpha$ morphologies in Sect.~\ref{sec:numSimu}. However, the two most frequent predicted morphologies are either corresponding to group {\bf G} (H$_\alpha$ emission aligned along the bar), or {\bf H}, when the emission is concentrated in the nucleus.  The configuration {\bf E}, where the star formation avoids the bar, is not reproduced in the simulation, in contradiction to our observations where the {\bf E} phase is about twice more frequent than the {\bf G} phase.

We think that this feature comes mainly from the star formation recipe adopted in the simulation. The efficiency of the star formation as a function of the gas density is still a challenging issue.  The predicted gas density appears to reproduce the observations. Our simulations reveal the formation of rings at the UHR, but this is always accompanied with a large gas density along the bar.

In most of the barred galaxies, the gas is observed in the bar, through the dust lane or through molecular line emission, like CO \citep[e.g.][]{2003ApJS..145..259H}, but little or no star formation and H$\alpha$ is observed. With the Schmidt law used in the present simulations (SFR $\propto \rho_{\rm{gas}}^{1.2}$), it is expected to find  recent star formation in the bar, while in the actual galaxies, this is avoided, maybe through too large relative velocities \citep[e.g.][]{1998A&A...337..671R}. Therefore to remove the discrepancy between H$_\alpha$ morphologies and predictions, it would be necessary to change the expression of the star formation rate: it should  not only depend on $\rho_g$ but should take into account the relative velocity of the gas with respect to the bar. This will be presented in a future work.

\begin{figure}
\includegraphics[angle=-90,width=\columnwidth]{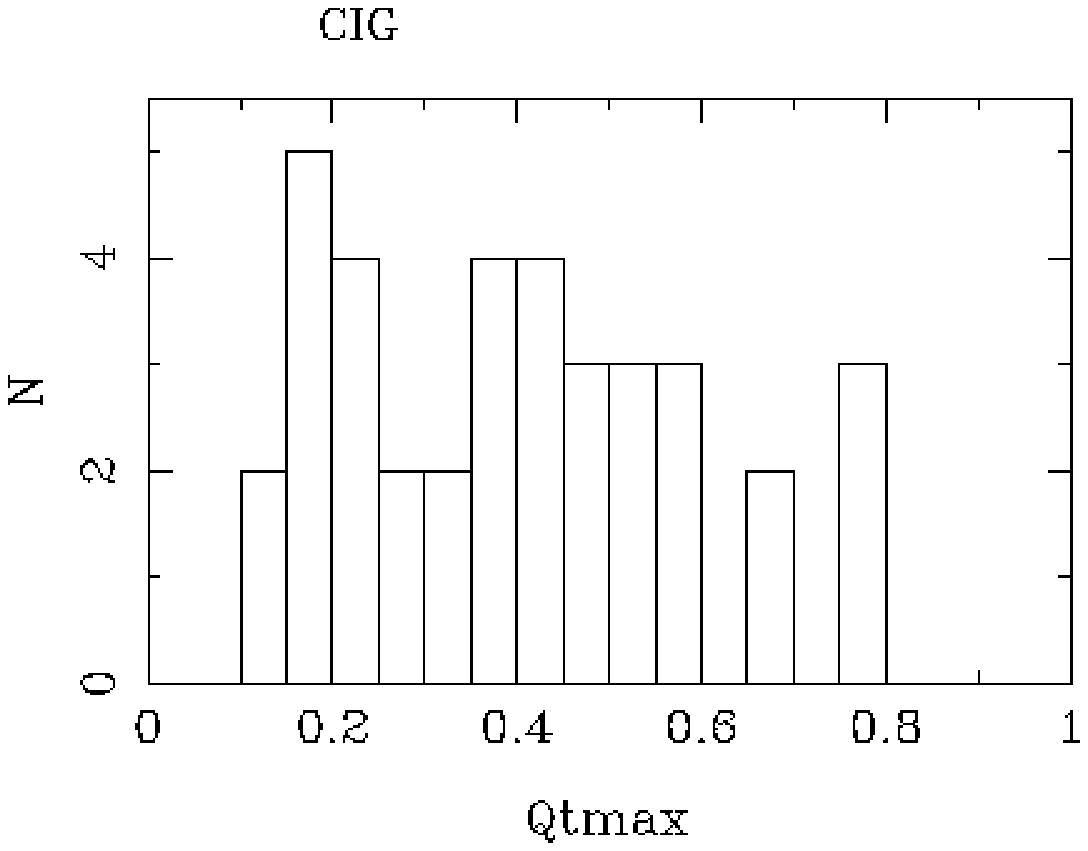} 
\caption{Distribution of total non-axisymmetric strengths in the CIG sample.} \label{fig:barStrength_CIG}
\end{figure}

\begin{figure}
\includegraphics[angle=-90,width=\columnwidth]{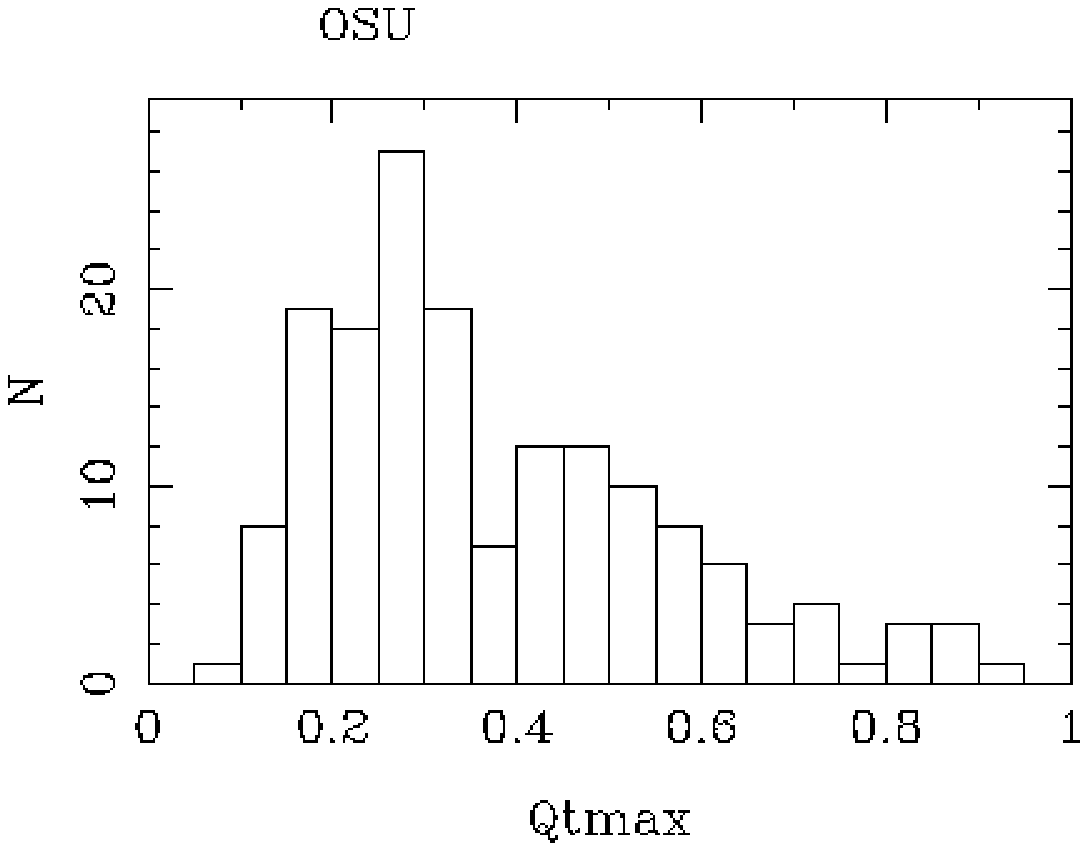}
 \caption{Distribution of total non-axisymmetric strengths in the OSU sample.} \label{fig:barStrength_OSU}
\end{figure}

The absence of star formation and H$_\alpha$ emission in the bar, while the gas density is abundant, is a frequently observed phenomenon. We do not have galaxies in common with the BIMA Survey Of Nearby Galaxies \citep[BIMA SONG,][]{2002ADIL...TH...01H}. But already, the BIMA SONG brings us some clues about the gas density. The {\bf E} phase, although also present there, is less frequent, relative to the {\bf G} phase, as we could expect. There is more CO emission in the bars of their galaxy sample (NGC 2903, 3627, 4535, 5457) compared to our H$_\alpha$ sample. The H$_\alpha$ emission is obviously  a more efficient tracer of star formation than of the total amount of gas available.

\subsubsection{Comparison with other samples}

Other groups have also imaged and classified spiral galaxies as a function of their H$_\alpha$ morphology. \citet{1999AJ....118..730H}, for instance, concentrated their work on 27 early-type spiral galaxies (Sa-Sab) in the nearby Universe. As our sample involves mainly Sb-Sc galaxies, the two samples are somehow complementary, although it has to be noted that their galaxies are selected without any regard to isolation criterion. We can group their galaxies according to our classification. The {\bf E} group would gather 11 galaxies: NGC\,1350, NGC\,1371, NGC\,1398, NGC\,1433, NGC\,1515, NGC\,1617, NGC\,2273, NGC\,3169, NGC\,5156, NGC\,7213 and 1108--48. Four galaxies would be classified in the {\bf G} group: NGC\,986, NGC\,5188, NGC\,7552 and NGC\,7582. The central emission characteristic of the {\bf H} group is seen in 5 galaxies: NGC\,1022, NGC\,1482, NGC\,3471, NGC\,3885, NGC\,5728. The galaxies NGC\,972, NGC\,5915 would be classified in the transitional {\bf EG} group while it would not be possible to classify five galaxies (NGC\,660, NGC\,2146, NGC\,3717, NGC\,6810, NGC\,7172) generally due to their too high inclination angle which makes difficult to disentangle the emission of the disk from the emission specific from the bar itself.

The comparison between the two samples shows that three groups present the same percentages of galaxies: the percentage of group {\bf E} galaxies is almost the same in the two samples (41 and 42\% in their and our sample, respectively). The same percentage of galaxies is also found for the {\bf G} group (15 and 18\%, respectively) and for the {\bf EG} group (about 7\% in both samples). Additionally, all their ENER (Extended nuclear emission-line region) galaxies fall into our {\bf E} group. Nevertheless, some differences are also observed. Our {\bf F} group is not represented in their sample, which lacks totally the galaxies without central emission. The other notable difference is that their sample presents a proportion 3 times higher of {\bf H} group galaxies. This can easily be explained because all their galaxies are Sa-Sab, hence more early-type (with less gas and very few H$_\alpha$ emission) than the bulk of our sample.\\

Other works targeted the inner H$_\alpha$ emission of galaxies: \citet{2006A&A...448..489K} focussed their work on the central regions (within 2~pc) of 73 spiral galaxies, that can have relatively close companions, which is the case for 26 galaxies (36\%) in their sample. We classified the 73 galaxies following our nomenclature: 45 galaxies (62\%) are in the {\bf E} group, 2 galaxies (3\%) in the {\bf F} group, 4 galaxies (5\%) in the {\bf G} group, 9 galaxies (12\%) in the {\bf H} group, 10 galaxies (14\%) in the {\bf EG} group, 1 galaxy (1\%) in the {\bf EF} group and 2 galaxies could not be classified ({\bf Irr} group), partly due to their to high inclination. The restriction of their sample only to the 26 galaxies wich possess a physically linked companion does not change significantly the percentages of the galaxies in the groups (61\% in {\bf E}, 4\% in {\bf F}, 0\% in {\bf G}, 19\% in {\bf H}, 12\% in {\bf EG}, 0\% in {\bf EF} and 1\% in {\bf Irr}). This is not surprising as \citet{2006A&A...448..489K} already noticed that the presence of a close companion does not produce any particular strong trend in the H$_\alpha$ morphology.

The very low percentage of {\bf F} galaxies with respect to our work is due to a selection criterion in the \citet{2006A&A...448..489K}'s sample as they selected galaxies with some prior evidence for H$_\alpha$ structure in their central regions. The same reason enhanced the proportion of group {\bf E} galaxies because they selected galaxies with already known nuclear or inner rings. Their sample is not complete and the comparison and interpretation with our work should be done with care. With respect to the isolation, the criterion they used can not ensure that the galaxies without ``very close companion'' would be isolated following Karachentseva's criterion (even if it seems that the presence of a companion does not drastically repercuss on the H$_\alpha$ features). As a result we cannot compare directly our isolated galaxies with their sample. Nevertheless, the \citet{1999AJ....118..730H} and \citet{2006A&A...448..489K} studies show that the groups we defined are also seen in samples that might not be isolated, so that the evolutive sequence could also be applied to galaxies marginally in interaction. For more detailed comparisons, samples better defined in term of isolation (or interaction) and better statistics would be needed.

\subsubsection{Bar frequency}

We have computed the total strength of bar and spiral waves in our galaxy sample by the method outlined in Sect.~\ref{sec:imAn}. We have adopted an exponential scaleheight equal to 1/12 of the radial scale length, independent of Hubble type, consistent with previous studies \citep[e.g.,][]{1994ApJ...437..162Q} although \citet{1998MNRAS.299..595D} found a type-dependent relationship and thicker disks. Varying this parameter will slightly change the absolute bar strengths in Figs.~\ref{fig:barStrength_CIG} \& \ref{fig:barStrength_OSU}, but will not have much impact on the relative values. The maximum of the parameter $Q_{\rm T}$, dominated by the bar strength and its harmonics, is used to build the histogram of the bar frequency for isolated galaxies (Fig.~\ref{fig:barStrength_CIG}). It is then possible to compare with the same histogram (see Fig.~\ref{fig:barStrength_OSU}) build for a general sample, the OSU sample \citep{2002ApJS..143...73E}. This sample of galaxies, selected without regard to any isolation criterion, appears to give quite similar results. Our histogram here is more uncertain and noisy, because of the much smaller sample size. However, it is possible to see that isolated galaxies sustain comparable bar strengths.

\section{Summary and conclusions} \label{sec:concl}

From the detailed morphological study of a sample of 45 isolated spiral galaxies, in H$_\alpha$ emission and optical red continuum, and comparison with numerical simulations, we come to the following results:

\begin{enumerate}
\item The percentage of bars (60\% in the optical) shows that isolated galaxies are not preferentially barred or unbarred galaxies. The histogram of relative tangential force strengths is quite similar to other samples of galaxies selected without regard to any isolation criterion.
\item Frequently, we observed a phase shift between gas and stellar components: the H$_\alpha$ emission is always leading with respect to the bar in the Gunn r image (shift angle $\sim10\degr$).
\item We interpreted the various global H$_\alpha$ morphologies observed in terms of the secular evolution experienced by galaxies in isolation. The main H$_\alpha$ classes can be related to the bar evolution phases. The observed frequency of particular patterns brings constraints on the time spent in the various evolution phases.
\item We chose to group the galaxies presenting similar features, we defined three main groups: group {\bf E} (19 galaxies) gathers the galaxies showing a strong central peak in the H$_\alpha$ emission, H$_\alpha$ emission along the spiral arms but not in the bar. Group {\bf F} (9 galaxies) is constituted by galaxies with less gas, which do not present any central emission knot in H$_\alpha$. Galaxies in group {\bf G} (8 galaxies) show H$_\alpha$ emission in the bar.
\item Other groups with fewer galaxies were also defined, most noticeable is the group {\bf EG} (3 galaxies) featuring galaxies having characteristics in between those defined by the groups {\bf E} and {\bf G}. Hence, we could interpret these features as different stages of an evolutive sequence: {\bf G} $\to$ {\bf EG} $\to$ {\bf E} $\to$ {\bf F}.
\item Numerical simulations showed a predicted frequency of the {\bf G} phase higher than the {\bf E} phase, in contradiction with our observations. We attribute this discrepancy to a failure of the star formation recipe, since we used the usual local Schmidt law for the star formation rate. The frequently observed phenomenon of star formation  avoiding the bar, in spite of large gas density there, suggests that the star formation law should depend also on other factors, in particular the relative velocity of the gas in the bar.
\end{enumerate}

\begin{acknowledgements}
We are very grateful to Gary A. Mamon, Chantal Balkowski, Alessandro Boselli, Santiago Garc\'\i a-Burillo and Jos\'e M. V\'\i lchez for their comments. We thank P. A. James for providing us data from the H$_\alpha$ Galaxy Survey and the anonymous referee for her/his helpful comments. The numerical simulations have been carried out on the Fujitsu NEC-SX5 of the CNRS computing center, at IDRIS (Palaiseau, France). We have made use of the LEDA database (http://leda.univ-lyon1.fr), and of the NASA/IPAC Extragalactic Database (NED) which is operated by the Jet Propulsion Laboratory, California Institute of Technology, under contract with the National Aeronautics and Space Administration. This work has been partially supported by DGI Grant AYA 2005-07516-C02-01 and the Junta de Andaluc\'\i a (Spain). GB is supported at the IAA/CSIC by an I3P contract (I3P-PC2005-F) funded by the European Social Fund.
\end{acknowledgements}


\end{document}